\documentclass[%
 reprint,
 amsmath,amssymb,
 pra,
]{revtex4-2}

\usepackage{graphicx}
\usepackage{dcolumn}
\usepackage{bm}

\usepackage{hyperref}
\begin{document}


\title{Broadband squeezed light field by magnetostriction in an opto-magnomechanical system}

\author{Ke Di\textsuperscript{1}, Shuai Tan\textsuperscript{1}, Anyu Cheng\textsuperscript{2}, Yinxue Zhao\textsuperscript{3}, Yu Liu\textsuperscript{1}, Jiajia Du\textsuperscript{1,*}}
\affiliation{\textsuperscript{1} Chongqing Engineering Research Center of Intelligent Sensing Technology and Microsystem, Chongqing University of Post and Telecommunications, Chongqing 400065, China\\
	\textsuperscript{2} Chongqing University of Post and Telecommunications, Chongqing 400065, China\\
	\textsuperscript{3} Wuhan Social Work Polytechnic, Wuhan, 430079, China\\
	}
\email{dujj@cqupt.edu.cn}

\date{\today}

\begin{abstract}
We present a novel mechanism for generating a wide bandwidth squeezed optical output field in an opto-magnomechanical system. In this system, the magnon (mechanical) mode in the yttrium-iron-garnet crystal is coupled to the microwave field (optical field) through magnetic dipole (radiation pressure) interaction. The magnetostrictive force induced by the yttrium-iron-garnet crystal causes a mechanical displacement and creates a quadrature squeezed magnon mode. Eventually, this quadrature squeezed mechanical mode is transferred to the output optical field through state-swap interaction. Our results demonstrate the optimal parameter range for obtaining a stable squeezed optical output field with a wide bandwidth. Moreover, the squeezed light field exhibits strong robustness to environmental temperature. The new scheme we propose has potential applications in quantum precision measurements, quantum wireless networks, quantum radar, etc.
\end{abstract}

\maketitle

\section{Introduction}
Squeezed light fields in continuous variables significantly enhance the detector sensitivity and calibrate the quantum efficiency of photodetection significantly\cite{2013Biological,PhysRevApplied.16.044031,PhysRevLett.117.110801}. Moreover, squeezed light fields find wide applications in quantum information processing\cite{PhysRevLett.94.053601,RevModPhys.84.621,WOS:000476554200001,2023Implementation}, quantum image enhancement\cite{PhysRevX.5.031004,casacio2021quantum}, and quantum precision measurement etc. As Einstein-Podolsky-Rosen (EPR) entanglement sources\cite{PhysRevD.23.1693,sudbeck2020demonstration,MA2022168357}, the preparation of squeezed light fields becomes the focus of attention. So far, traditional methods for preparing squeezed light fields include parametric down conversion of nonlinear crystals (optical parametric amplification)\cite{PhysRevLett.117.110801,PhysRevLett.57.2520,PhysRevLett.93.161105,PhysRevLett.101.233602,PhysRevLett.106.153602,2022Phase}, four-wave mixing \cite{PhysRevLett.57.691,PhysRevLett.55.2409,Sharping:01,guo2012all,PhysRevLett.124.193601,WOS:000963167600001}, and optomechanical systems\cite{2013Strong,PhysRevLett.118.103601,Xiong:22} etc. In recent years, new schemes for generating squeezed light fields have emerged, such as photonic chips based on microresonators and remote state preparation\cite{yang2021squeezed,Han:22}. While these schemes take different forms, most of them still rely on the nonlinear physical processes of light \cite{andersen201630}.

In this paper, we present a new mechanism based on an opto-magnomechanical system, which distinguishes it from all previous approaches for preparing continuous variable squeezed light fields. The opto-magnomechanical system consists of a micro-bridge structured yttrium-iron-garnet (YIG) crystal. It acts as a connector, linking a microwave cavity and an optical cavity on opposite sides, respectively. The magnon mode in this crystal is coupled to the deformed phonon mode by magnetostrictive forces\cite{PhysRevLett.121.203601,li2023squeezing}. The radiation pressure-like dispersive interactions between the magnon and phonon provide the necessary nonlinearity for creating squeezed spin waves\cite{li2023squeezing}. The magnetostrictive effect induces a correlation between the amplitude and phase of the mechanical mode. This correlation gives rise to a mechanical quadrature squeezing. Furthermore, the optical cavity mode is coupled to geometrical deformation of YIG crystal via radiation pressure in optomechanical\cite{PhysRevA.49.4055,Di:23}. We activate opto-mechanical anti-Skotos scattering by driving the optical cavity with a red detuned laser field\cite{fan2022stationary}. An effective mechanical-optical state-swap (beamsplitter) interaction is established. Therefore, the quadrature mechanical squeezing state created by the magnetostrictive force is available for transfer to the optical field via state-swap interaction. The output optical field of the squeezed state is eventually detected by a homodyne detection system\cite{PhysRevLett.106.153602}.

\section{Theoretical Model}
\begin{figure}[ht]
	\centering
	\includegraphics[scale=0.18]{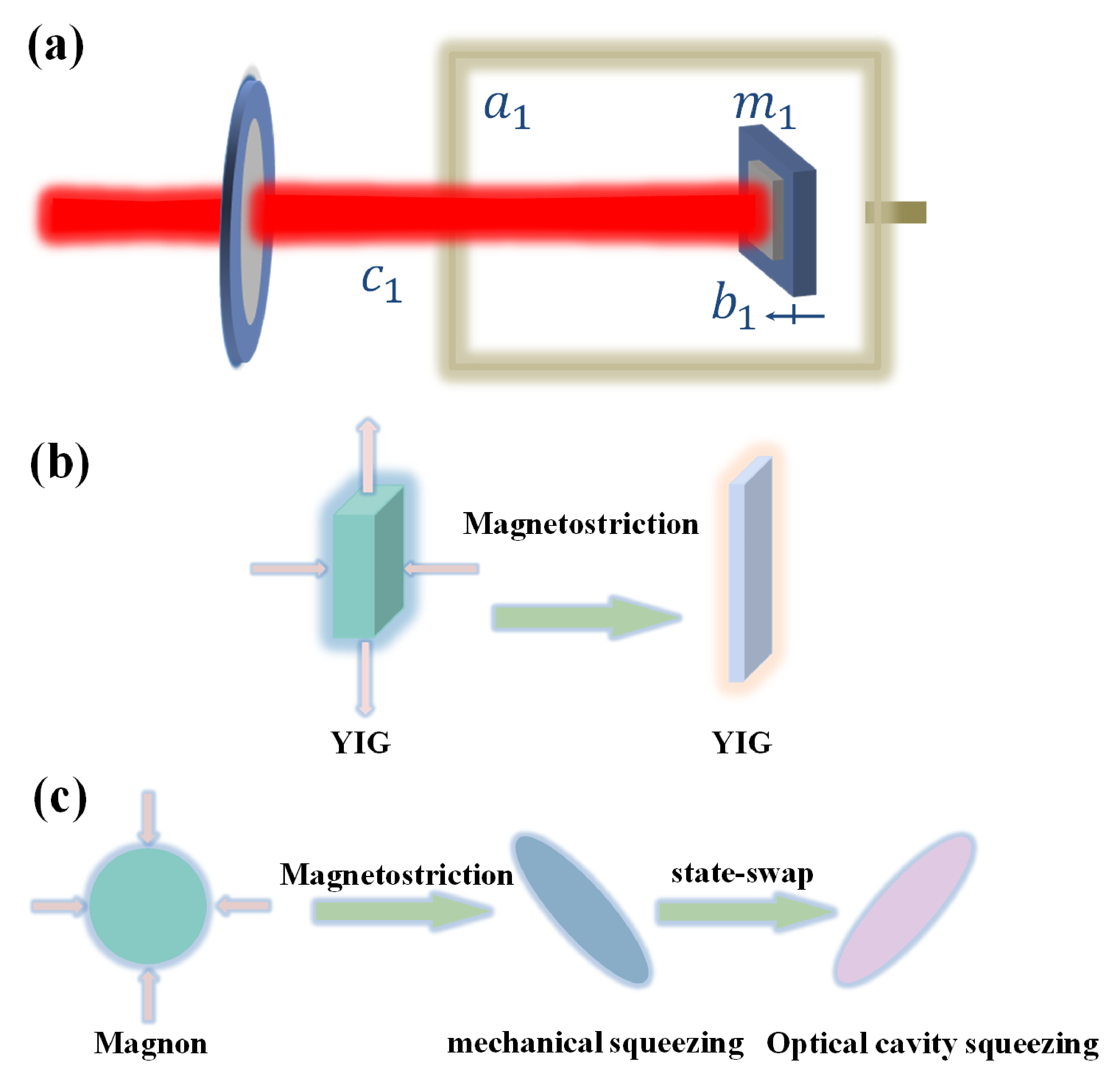}
	\caption{Scheme diagram of the opto-magnomechanical system. (a) Model diagram of the system to realize a broadband squeezed light field. The YIG micro-bridges are embedded in a microwave cavity. High-reflection mirrors are attached to the right side of the YIG crystal, which are used to construct the optical cavity. The laser and microwave fields are injected from the left and right sides of the system model diagram, respectively. (b) The process of YIG crystal displacement deformation induced by magnetostrictive forces at the macroscopic level. (c) Mechanism for generating squeezed light field. The magnetostrictive forces induce mechanical mode to generate squeezed state. This squeezed state is transferred to the optical cavity field via state-swap interaction.}
	\label{fig:1}
\end{figure}

The opto-magnomechanical system, illustrated in Fig. \ref{fig:1}, encompasses a microwave cavity mode, a magnon mode, a phonon mode, and an optical cavity mode. The collective spin excitation in YIG crystal $\left( {\rm a}\; 5 \times 3 \times 2\; \mu m^{3}\; {\rm YIG\; cuboid} \right)$ with micro-bridge is described by the magnon mode \cite{fan2022stationary,fan2022microwave}. The phonon mode describes the mechanical vibration of a YIG crystal which is excited by the uniform bias magnetic field and microwave cavity. Additionally, the optical cavity in this system comprises of two highly reflective mirrors. The right mirror is positioned on the surface of the YIG micro-bridge. The coupling between the magnon mode and the microwave cavity mode is accomplished through the magnetic dipole coupling\cite{PhysRevLett.113.083603}. The dispersive coupling between the magnons and the low-frequency vibrational phonons is achieved by the magnetostrictive interactions\cite{PhysRevLett.121.203601}.  Furthermore, the phonon mode couples with the optical cavity mode through radiation pressure\cite{Di:23,RevModPhys.86.1391}. The dispersive coupling between magnons and phonons is essential which provides the nonlinearity required to create the squeezed state in an opto-magnomechanical system\cite{li2023squeezing}. The radiation pressure interaction between the phonon mode and the optical cavity mode enables state-swap (beamsplitter) interaction\cite{Di:23}. Therefore, the squeezed state of the magnon can be transferred to the optical field.

The Hamiltonian representing the cavity optomagnomechanics system can be expressed as \cite{PhysRevLett.121.203601,yu2020macroscopic}:
\begin{equation}
	\begin{aligned}
		H/\hbar=&\omega_{a}a^{+}a+\omega_{m}m^{+}m+\omega_{c}c^{+}c+\frac{\omega_{b}}{2}(q^2+p^2)\\
		&+g_{ma}(a^{+}m+am^{+})+g_{mb}m^{+}mq-g_{bc} c^{+}cq\\
		&+i\Omega(m^{+}e^{-i\omega_{o}t}-me^{i\omega_{o}t})+iE(c^{+}e^{-i\omega_{L}t}-ce^{i\omega_{L}t}).
		\label{eq:1}
	\end{aligned}
\end{equation}
where $a$, $m$ and $c$ ($a^+$, $m^+$ and $c^+$) are the annihilation (creation) operators for the microwave cavity mode, the magnon mode, and the optical cavity mode, respectively. Each mode satisfies $[O,O^{+}]=1$ ($O=a, m, c $). $q$ and $p$ are the dimensionless position and momentum operators for the mechanical vibration mode. It  satisfies $[q,p]=i$. $\omega_j$ ($j=a,m,c,b$) denote the resonant frequencies of the microwave cavity, magnon, optical cavity, and phonon mode, respectively. The first four terms in Eq.(\ref{eq:1}) represent the energy contributions of different modes\cite{Di:23,fan2022microwave}. $g_{ma}$ is the coupling strength of the magnetic dipole between the microwave cavity mode and the magnon mode. The $g_{mb}$ and $g_{bc}$ denote the magnon-phonon bare coupling rate and the phonon-optical cavity bare coupling rate. The fifth, sixth and seventh terms in Eq.(\ref{eq:1}) are the interactions between different modes. The eighth and ninth terms describe the driving fields for the magnon mode and the optical cavity mode respectively\cite{PhysRevLett.121.203601,Di:23}. In these terms, the Rabi frequency is denoted as $\Omega=\frac{\sqrt{5}}{4}\gamma \sqrt{N_{s}}H_{d}$, while the coupling strength of the optical cavity is represented as $E=\sqrt{2\kappa_{c}P_{L}/(\hbar\omega_{L})}$\cite{PhysRevLett.121.203601,fan2022microwave}. $\gamma$, $N_{s}$ and $H_{d}$ denote the gyromagnetic ratio, the total spin number of the ferrimagnet and the amplitude of the drive magnetic field. $\kappa_{c}$ describe the decay rate of the optical cavity mode.  $P_{L}$ and $\omega_{L}$ are the power and frequency of the laser with wavelength of 1550 nm. 

According to the equation $\dot{O}=\frac{1}{i\hbar}[O,H]-\frac{\kappa_{O}}{2}O+\sqrt{\kappa_{O}}O^{in}$, we can derive the quantum Langevin equation for different modes as follows \cite{PhysRevLett.121.203601,vidal2002computable}:
\begin{equation}
	\begin{aligned}
		\dot{a}
		=&-i\Delta_{a}a-\frac{\kappa_{a}}{2}a-ig_{ma}m+\sqrt{\kappa_{a}}a_{in},\\
		\dot{m}
		=&-i\Delta_{m}m-\frac{\kappa_{m}}{2}m-ig_{ma}a-ig_{mb}mq+\Omega\\&+\sqrt{\kappa_{m}}m_{in},\\
		\dot{c}	=&-i\Delta_{c}c-\frac{\kappa_{c}}{2}c+ig_{bc}cq+E+\sqrt{\kappa_{1}}{c^{in}_1}+\sqrt{\kappa_{2}}{c^{in}_2},\\
		\dot{p}
		=&-\omega_{b}q-{\gamma_b}p+{g_{bc}}{c^+}c-{g_{mb}}{m^+}m+\xi,\\
		\dot{q}
		=&\omega_{b}p, 	
		\label{eq:2}
	\end{aligned}
\end{equation}
In a rotating frame with frequencies $\omega_{0}$ and $\omega_{L}$, $\Delta_n=\omega_{n}-\omega_{0}$ ($n=a,m$),  $\Delta_{c}=\omega_{c}-\omega_{L}$. $\kappa_{a}(\kappa_{m})$ is decay rate of the microwave cavity (magnon) mode, $\gamma_b$ describes the mechanical damping rate. $\xi$ is a Brownian stochastic force that can be assumed to be Malkovian since a large mechanical quality factor $Q = \omega_{b}/\gamma_b\gg1$\cite{PhysRevLett.46.1}. The correlation function: $\langle\xi(t)\xi(t^{'})+\xi(t^{'})\xi(t)\rangle/2 \simeq \gamma_b[2N_{b}(\omega_{b})+1] \delta(t-t^{'})$. $a^{in}$ and $m^{in}$describe the input noise operators for the microwave cavity mode and the magnon mode, respectively. $c^{in}_1$ and $c^{in}_2$ denote the input noise entering the optical cavity via the left-end mirror and the input noise of all other decay channels, respectively. The total decay rate $\kappa_{c} \equiv \kappa_{1}+\kappa_{2}$. The zero-mean correlation function $O^{in}$ of the input noise can be described as \cite{PhysRevLett.106.220502,PhysRevApplied.9.044023} $\langle O^{in}(t)O^{in+}(t^{'}) \rangle = [N_{O}(\omega_{O})+1] \delta(t-t^{'})$, and $\langle O^{in+}(t)O^{in}(t^{'}) \rangle = N_{O}(\omega_{O}) \delta(t-t^{'})$ ($O=a, m, c $).  Here, $ N_{j}(\omega_{j}) = [exp(\frac{\hbar \omega_{j}}{k_{B}T})-1]^{-1}$ $(j=a, m, c, b)$ with the Boltzmann constant $k_{B}$ and  the environmental temperature $T$.

To create a stationary opto-magnomechanical system that we employ a strong drive for the magnon and optical cavity modes. Therefore, large coherent magnon mode and optical mode amplitudes $|\langle m \rangle|\gg1$  and $|\langle c \rangle|\gg1$ are obtained at the steady state\cite{li2021entangling}.  This allows us to linearize the system dynamics around large mean values.  In the form of a fluctuation around a prominent average value, each operator can be described, i.e., $O=\langle O\rangle+\delta O$ ($O=a,m,c$). Therefore, the fluctuation term of the system Langevin equation of motion is given by
\begin{equation}
	\begin{aligned}
		\delta\dot{a}
		=&-i\Delta_{a}\delta a-\frac{\kappa_{a}}{2}\delta a-ig_{ma}\delta m+\sqrt{\kappa_{a}}a_{in},\\
		\delta\dot{m}
		=&-i\widetilde{\Delta}_{m} \delta m-\frac{\kappa_{m}}{2}\delta m-ig_{ma} \delta a-iG_{mb} \delta q\\&+\sqrt{\kappa_{m}}m_{in},\\
		\delta\dot{c}	
		=&-i\widetilde{\Delta}_{c}\delta c-\frac{\kappa_{c}}{2} \delta c+iG_{bc} \delta q+\sqrt{\kappa_{1}}{c^{in}_1}+\sqrt{\kappa_{2}}{c^{in}_2},\\
		\delta\dot{p}
		=&-\omega_{b}\delta q-{\gamma_b} \delta p-G_{mb}^{*} \delta m-G_{mb}\delta m^{+}+G_{bc}^{*} \delta c \\&+ G_{bc} \delta c^{+}+\xi,\\
		\delta\dot{q}
		=&\omega_{b} \delta p, 	
		\label{eq:3}
	\end{aligned}
\end{equation}
where $\widetilde{\Delta}_{m}=\Delta_{m} + g_{mb} \langle q \rangle$ and $\widetilde{\Delta}_{c}=\Delta_{c} - g_{bc} \langle q \rangle$  are the effective detuning containing the frequency shift terms in the system. Mechanical displacement is usually small, 
$|\widetilde{\Delta}_{m}|\simeq|\Delta_{m}|, |\widetilde{\Delta}_{c}|\simeq|\Delta_{c}|$, $|\Delta_{a}|, |\widetilde{\Delta}_{m}|, |\widetilde{\Delta}_{c}|\simeq\omega_{b}$.
$ \langle q \rangle=(g_{bc}|\langle c \rangle|^{2}-g_{mb}|\langle m \rangle|^{2}|)/\omega_{b}$ denotes the steady-state displacement. The steady-state amplitudes $\langle m \rangle$ and $\langle c \rangle$ are
\begin{equation}
	\begin{aligned}
		\langle m \rangle =&\frac{ \Omega (\kappa_{a}/2+i \Delta_a)}{g^{2}_{a}+(\kappa_{m}/2+i\widetilde{\Delta}_{m})(\kappa_{a}/2+i\Delta_{a})},\\
		\langle c\rangle
		=&\frac{E}{\kappa_{c}/2+i\widetilde{\Delta}_{c}},
		\label{eq:4}
	\end{aligned}
\end{equation}
$G_{mb}=g_{mb} \langle m \rangle$ and $G_{bc}= g_{bc} \langle c \rangle$ describe effective coupling  rates of magnon-phonon and phonon-optical cavity, respectively. The drift matrix can be extracted from Eq.(\ref{eq:3}), which can be described in matrix form $\dot{u}(t)=Au(t)+n(t)$. Where $u(t)=(\delta a, \delta a^{+},\delta m,\delta m^{+}, \delta c, \delta c^{+}, \delta q, \delta p)^T$, $n(t)=(\sqrt{\kappa_{a}}a^{in}, \sqrt{\kappa_{a}}a^{in+}, \sqrt{\kappa_{m}}m^{in}, \sqrt{\kappa_{m}}m^{in+}, \sqrt{\kappa_{1}}c^{in}_{1}+\sqrt{\kappa_{2}}c^{in}_{2}, \sqrt{\kappa_{1}}c^{in+}_{1}+\sqrt{\kappa_{2}}c^{in+}_{2}, 0 , \xi)^T$.  A is the drift matrix:
\begin{equation}
	\begin{aligned}
		A=\left(\begin{array} {c c c c }	
			A_{a}  & A_{am} &  A_{ab} &  A_{ac}\\
			A_{ma} & A_{m}  & A_{mb} &  A_{mc}\\		
			A_{ca} & A_{cm} &  A_{c} &  A_{cb}	\\
			A_{ba} & A_{bm} &  A_{bc}  & A_{b}\\				
		\end{array}\right),
		\label{eq:5}			
	\end{aligned}
\end{equation}
where $A_o=diag(-i\Delta_o-\frac{\kappa_o}{2}, i\Delta_o-\frac{\kappa_o}{2})$, ($ o = a, m, c$), $A_b = (0, \omega_b; -\omega_b, -\gamma_b)$. $A_{am}=A_{ma}=diag(-ig_{ma},ig_{ma})$, $A_{mc}=(-iG_{mb},0;iG_{mb},0)$, $A_{cb}=(iG_{bc},0;-iG_{bc},0)$, $A_{bm}=(0,0;-G_{mb}^{*},-G_{mb}^{*})$ and $A_{bc}=(0,0;G_{bc}^{*},G_{bc}^{*})$. The other matrix blocks are the $2\times2$ zero matrices. The eigenvalues of the drift matrix $A$ obtained from the data of this scheme are all negative real numbers, thus the system works in steady-state \cite{dejesus1987routh,kong2022magnon}.

Taking the Fourier transform for Eq.(\ref{eq:3}), we can obtain the expression of the Langevin motion equation in the frequency domain. The expression for the quantum fluctuations in the frequency domain takes the following form \cite{li2023squeezing}:
\begin{equation}
	\begin{aligned}
		\delta F(\omega)=&F_{A}(\omega) a^{in}(\omega)+F_{a}(\omega) a^{in+}(-\omega)+
		F_{M}(\omega) m^{in}(\omega)\\&+F_{m}(\omega) m^{in+}(-\omega)
		+\sum_{j=1,2}[F_{C_j}c^{in}_{j}(\omega)\\&+F_{c_{j}}c^{in+}_{j}(-\omega)]+F_{B}(\omega)\xi(\omega),
		\label{eq:6}
	\end{aligned}
\end{equation}
where $F_{k}(\omega)$ $(k=A,a,M,m,C,c,B)$ are the input noise coefficient of different modes. Furthermore, this allows us to study the noise properties of the reflected optical field based on the standard input-output relationship. In order to detect conveniently, we define the quadrature of the optical cavity output field as
\begin{equation}
	\begin{aligned}
		\delta W^{o}(\omega)=\frac{1}{\sqrt{2}}[\delta c^{o}(\omega)e^{-i\phi}+\delta c^{o+}(-\omega)e^{i\phi}],
		\label{eq:7}
	\end{aligned}
\end{equation}
where $\phi$ is the phase angle. Quantum fluctuation of the output field $\delta c^{o}(\omega)$ is obtained via the input-output relation $\delta c^{o}(\omega)=\sqrt{\kappa_{1}}\delta c(\omega)-\delta c^{in}_{1}(\omega)$. The quadrature of the noise spectral density for the optical cavity output field is defined as \cite{li2023squeezing} 
\begin{equation}
	\begin{aligned}
		S^{o}(\omega)=&\frac{1}{4\pi}\int_{-\infty}^{+\infty}d\omega^{'}e^{-i(\omega+\omega^{'})t} \times \langle \delta W^{o}(\omega) \delta W^{o}(\omega^{'})\\&+\delta W^{o}(\omega^{'}) \delta W^{o}(\omega) \rangle,
		\label{eq:8}
	\end{aligned}
\end{equation}
Eq.(\ref{eq:8}) is the criterion for identifying whether the output field is in the squeezed state. If $S^{o}(\omega)<\frac{1}{2}$, it denotes that the output optical field is squeezed.

\section{Numerical results of squeezed optical}

\begin{figure}[ht]
	\centering
	\includegraphics[scale=0.15]{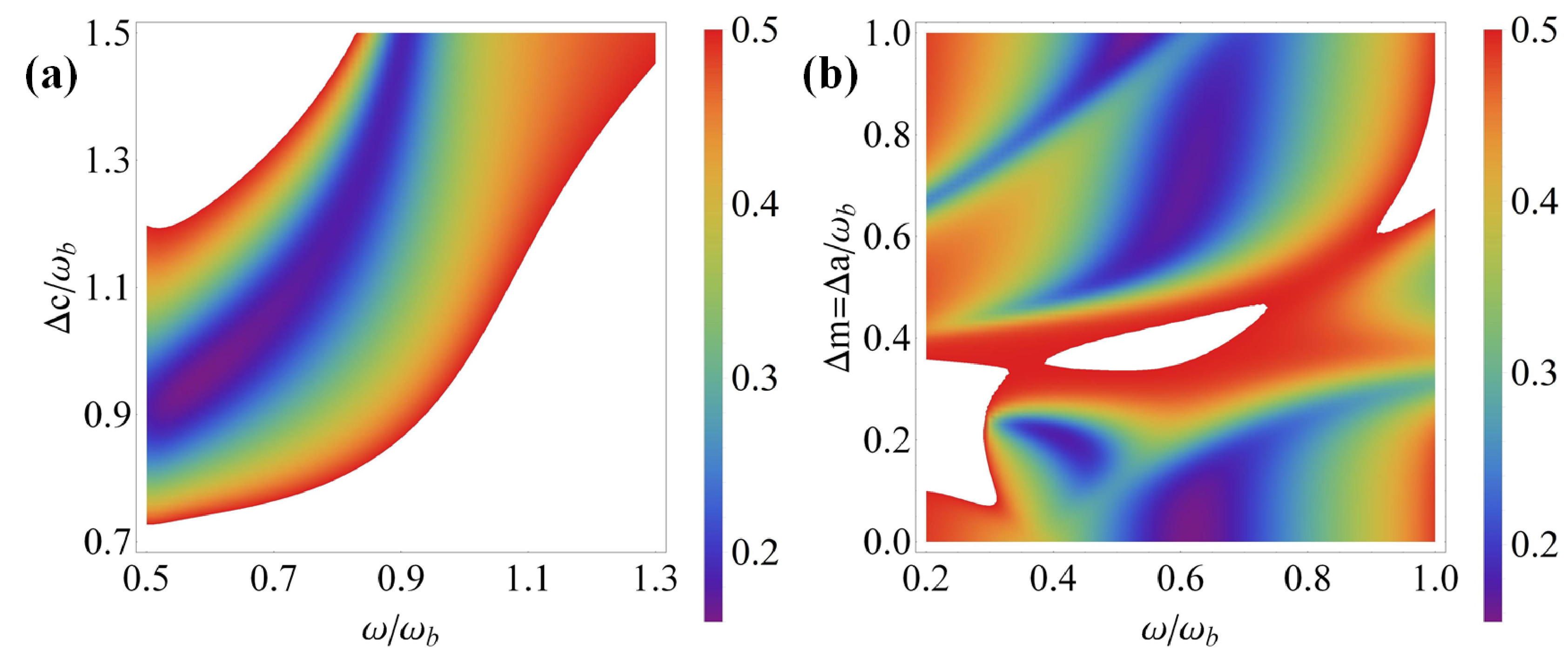}
	\caption{Noise spectral density $S^{o}(\omega)$ versus detuning $\Delta$ and frequency $\omega$. (a) Noise spectral density $S^{o}(\omega)$ of output optical cavity mode versus the detuning $\Delta_c$ and frequency $\omega$ with $\Delta_{m}=\Delta_{a}=0.1\omega_b$, $\phi=0.3\pi$, $\kappa_{c}=\omega_b$. The colored column at the right side indicates strength of noise spectral density $S^{o}(\omega)$. The blank area in the figure denotes the noise spectral density higher than the shot-noise level ($S^{o}(\omega)>0.5$). (b) Noise spectral density $S^{o}(\omega)$ of output optical cavity mode versus the detuning $\Delta_m=\Delta_a$ and frequency $\omega$ with $\Delta_{c}=\omega_b$. 
	}
	\label{fig:2}
\end{figure}

Fig. \ref{fig:2} shows the effect of different detuning modes ($\Delta_{c},\Delta_{m}$ and $\Delta_{m}$) versus frequency $\omega$ for the noise spectral density $S^{o}(\omega)$ of the optical cavity output field, respectively. 
We demonstrate that a squeezed light field with fluctuations below the shot-noise level can be created by an  opto-magnomechanical system using experimentally feasible parameters. The feasibility parameters adopted experimentally in this paper\cite{PhysRevLett.124.213604,zhang2016cavity,fan2022microwave,PhysRevApplied.12.054031,PhysRevX.11.031053}: $\omega_{a}=\omega_{m}=2\pi \times 10\;{\rm GHz}$, $\omega_{b}=2\pi \times 40\;{\rm MHz},\lambda_{c}=1550\;{\rm nm},P_L=0.64\;{\rm mW}, \kappa_{a}=2\pi \times 5\;{\rm MHz},\kappa_{m}=2\pi \times 2\;{\rm MHz}, \kappa_{2}=0.1\omega_{b}, \gamma_{b}=2\pi \times100\;{\rm Hz}, g_{mb}=2\pi \times20\;{\rm Hz},g_{bc}=2\pi \times4\;{\rm kHz}, g_{ma}=2\pi \times15\;{\rm MHz}$ and $ T=20\;{\rm mK} $. The strength of the driving magnetic field is $H_{d}=\sqrt{2\mu_0 P_0/(l w c)}$, where the power is $P_0\simeq 5$ mW and the vacuum magnetic permeability is $\mu_0=4\pi \times 10^{-7}$. $l=5$ $\mu m$ ($w=3$ $\mu m$) describes the length (width) of the YIG micro-bridge \cite{fan2022microwave}. The squeezing in the opto-magnomechanical system is generated by magnetostrictive forces acting on the mechanical modes, which exhibit correlated switching in both amplitude and phase. Additionally, we activate the optical mechanical anti-Skotos scattering by driving the optical cavity with a red-detuned laser field. This creates a beamsplitter-type (state-swap) coupling between phonon and optical field, allowing the squeezed state generated by the magnetostrictive forces to be transferred to the optical field. In order to obtain a strong magnetostrictive effect, the magnon frequency is always tuned to resonate with the microwave cavity ($\Delta_{a}=\Delta_{m}$). In Fig. \ref{fig:2}(a), the stationary squeezed light field appears around frequency $\omega \simeq0.65\omega_b$ and detuning $\Delta_{c} \simeq \omega_b$. The minimum noise spectral density $S^{o}(\omega)$ is 0.16 (corresponding to 4.95 dB below the vacuum fluctuation). The stationary squeezed state output light field is detected in the frequency range of $0.5\omega_b \sim 0.9\omega_b$. The widest frequency band is up to $2\pi\times16$ MHz.  Moreover, the effect of the magnon and microwave field resonant frequency on the noise spectral density is investigated at the optimal optical cavity frequency. The results shown in Fig. \ref{fig:2}(b) demonstrate that the magnon and microwave field resonant frequency is able to create a stationary squeezed light field at several different frequencies. Considering the convenience of experimental implementation, tuning the frequency of the optical field and holding the resonant frequency of the magnon and microwave field ($\Delta_{m}=\Delta_{a}=0.1\omega_b$). This enables to obtain more stable and wider bandwidth squeezed states as shown in Fig. \ref{fig:2}(a). Maximum squeeze is also obtained at $\Delta_{m}=\Delta_{a}=0.1\omega_b$.
\begin{figure}[h]
	\centering
	\includegraphics[scale=0.15]{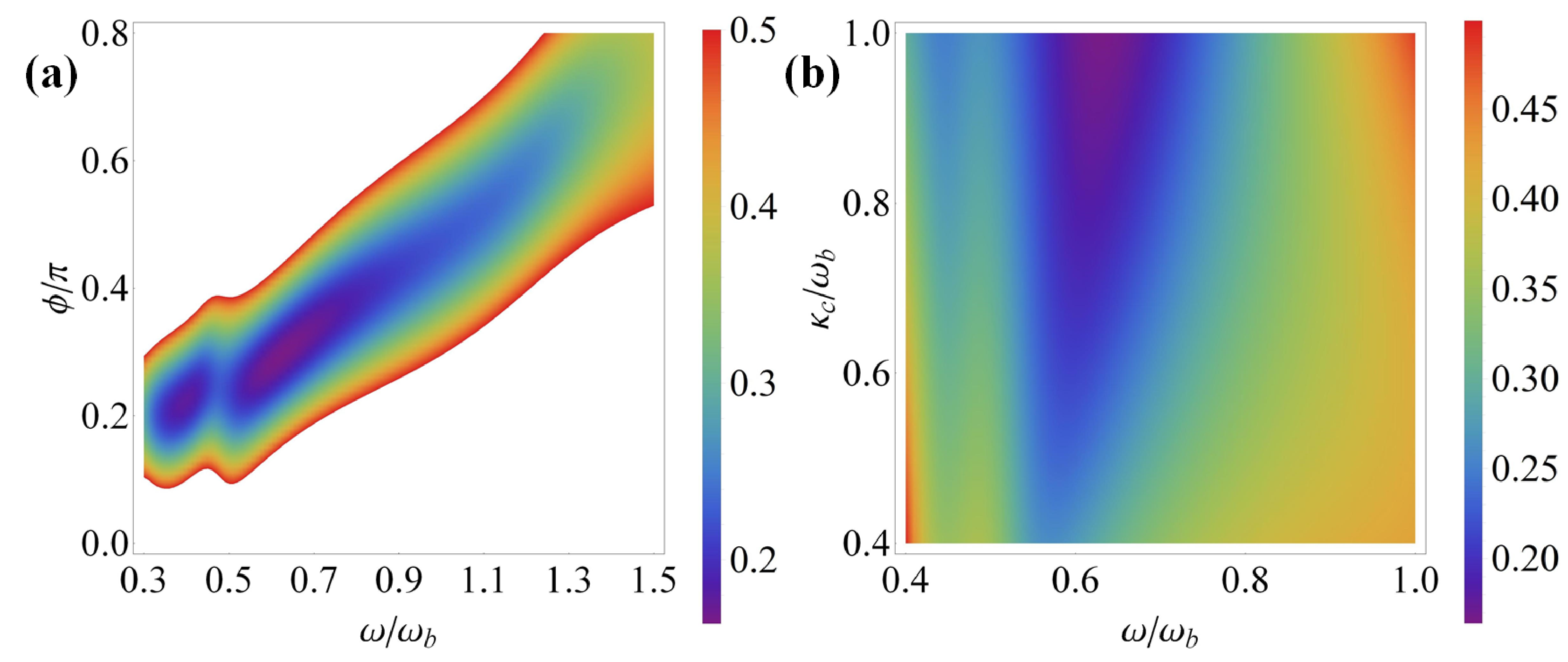}
	\caption{Noise spectral density $S^{o}(\omega)$ with respect to phase $\phi$ and the decay rate $\kappa_{c}$ of the optical cavity mode. (a) Noise spectral density $S^{o}(\omega)$ of output optical cavity mode versus the phase $\phi$ and frequency $\omega$ with $\Delta_{m}=\Delta_{a}=0.1\omega_b$, $\Delta_{c}\simeq\omega_b$, $\kappa_{c}=\omega_b$. (b) Noise spectral density $S^{o}(\omega)$ of output optical cavity mode versus the decay rate $\kappa_{c}$ and frequency $\omega$ with $\phi=0.3\pi$. Other parameters are the same as Fig. \ref{fig:2}.
	}
	\label{fig:3}
\end{figure}

Next, we describe the noise spectral density $S^{o}(\omega)$ of the output optical field versus phase $\phi$ and frequency $\omega$ in Fig. \ref{fig:3}(a). The results show that the phase $\phi=0.3\pi$ creates the optimal optical field squeezed state. The corresponding minimum noise spectral density $S^{o}(\omega)$ is 0.17 (4.68 dB below vacuum fluctuation). As the phase gradually increases, the noise spectral density $S^{o}(\omega)$ increases corresponding to the decrease in the squeezed degree of the output optical field until it disappears. Furthermore, Fig. \ref{fig:3}(b) explores the effect of the optical cavity decay rate $\kappa_{c}$ on the noise spectral density $S^{o}(\omega)$. A large optical cavity decay rate is available for the reflected optical field to be output from the optical cavity. Therefore, the output field of the optical cavity obtains a stable and broadband squeezed light field when the decay rate $\kappa_{c} $ of the optical cavity gradually increases to $ \omega_b$.

\begin{figure}[htbp]
	\centering
	\includegraphics[scale=0.13]{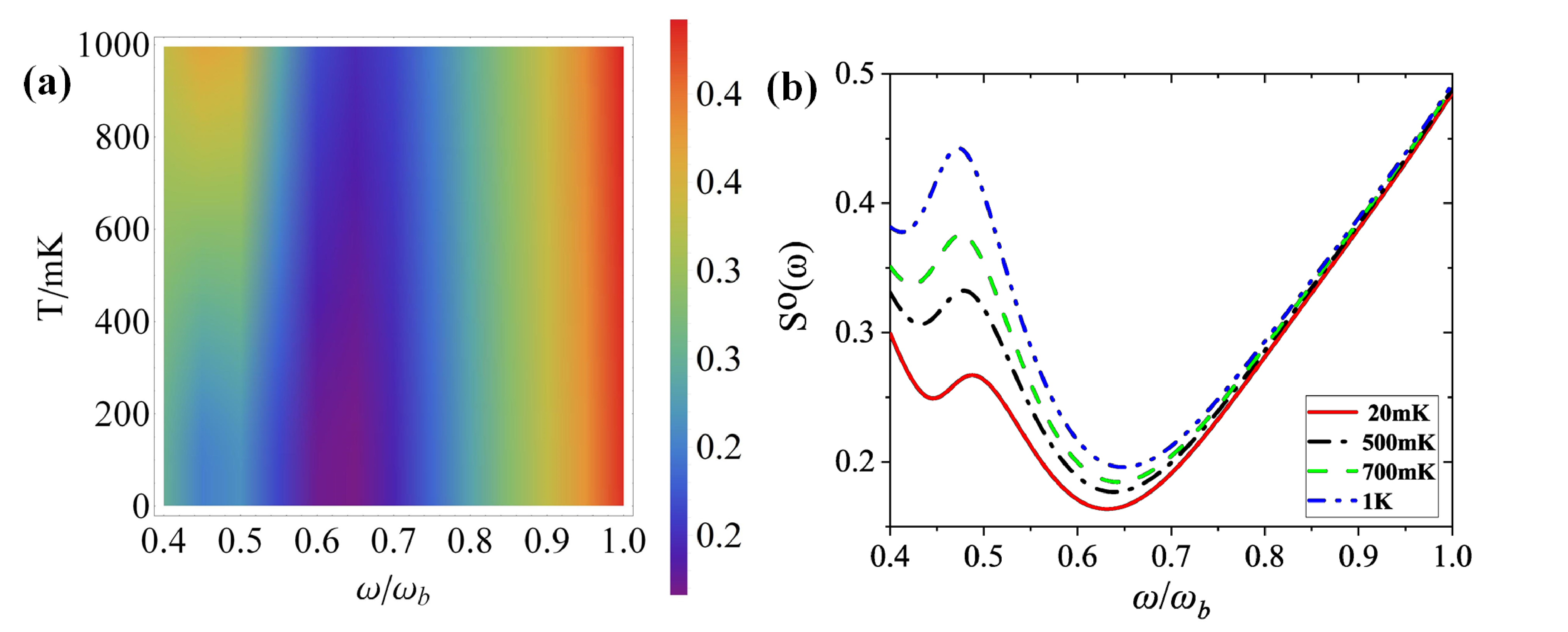}
	\caption{Noise spectral density $S^{o}(\omega)$ versus environmental temperature $T$. (a) Noise spectral density $S^{o}(\omega)$ of output optical cavity mode versus the environmental temperature $T$ and frequency $\omega$ with $\Delta_{m}=\Delta_{a}=0.1\omega_b$, $\Delta_{c}\simeq\omega_b$, $\kappa_{c}=\omega_b$, $\phi=0.3\pi$. (b) Noise spectral density $S^{o}(\omega)$ of output optical cavity mode at various temperatures: $T=20$ mK, $T=500$ mK,  $T=700$ mK,  $T=1$ K. Other parameters are the same as Fig. \ref{fig:2}.
	}
	\label{fig:4}
\end{figure}

In Fig. \ref{fig:2} and \ref{fig:3}, the output squeezed light field is in an ultra-low temperature $T=20$ mK environment. In Fig. \ref{fig:4}, we investigate the effect of environmental temperature on the noise spectral density. The results of this research indicate that the squeezed light field created by the opto-magnomechanical system is strongly robust to thermal noise. The best robustness to environmental temperature is found around frequency $\omega \simeq 0.65 \omega_b$. The noise spectral density $S^{o}(\omega)$ of the output field increases from 0.16 to 0.19 (4.95 dB to 4.2 dB below vacuum fluctuation) as the environmental temperature $T$ increases from 20 mK to 1 K. Significant variation in environment temperature has a weak effect on the output squeezed optical field.

Finally, we discuss a possible experimental scheme for detecting the squeezed-state optical field. This experimental scheme consists of the opto-magnomagnetic system, mode clean system, and homodyne detector system\cite{PhysRevLett.117.110801, PhysRevLett.106.153602}. The 1550 nm beam which passes through the mode clean cavity is split by a polarizing beam splitter (PBS) into local oscillation light and driving laser. Additionally, the local oscillation light and the squeezed output light field are simultaneously directed through a $50/50$ beam splitter. The quadrature of the squeezed light field can be measured by homodyning the field which passes through a $50/50$ beam splitter.

\begin{figure}[htbp]
	\centering
	\includegraphics[scale=0.4]{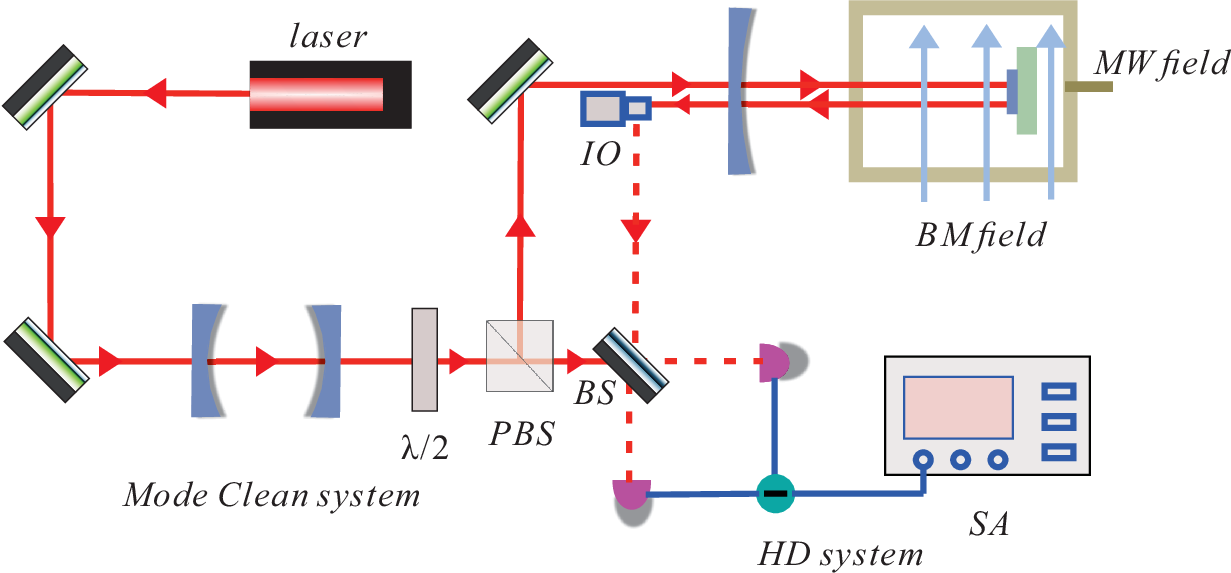}
	\caption{Diagram of a feasible experimental scheme. OI: optical isolator, MW field: microwave field, BM field: bias magnetic field, BS: beam splitter, PBS: polarizing beam splitter, HD: homodyne detector,  SA: spectrum analyzer.
	}
	\label{fig:5}
\end{figure}

\section{Conclusion}
We present a novel mechanism based on magnetostrictive effects to create a continuous variable squeezed light field in an opto-magnomechanical system. This mechanism leverages the magnetostrictive effect and radiation pressure interaction in optomechanical theory to generate a stable squeezed optical output field with a wide bandwidth. In this paper, we analyze in detail the impact of different mode detuning, phase, decay rate, and environmrntal temperature of the optical cavity on the noise spectral density. Ultimately, we determine the optimal range of parameters that enables the generation of a stationary squeezed light field and discover the robustness of the scheme to environmental thermal noise. Additionally, we propose an experimentally implementable detection scheme. Our work may find applications in quantum precision measurements\cite{PhysRevA.106.013506,PhysRevLett.130.073601}, quantum wireless networks\cite{shi2021entanglement,PhysRevApplied.18.044002}, etc.

\section*{Funding}
This work was supported by National Natural Science Foundation of China (Grant Nos.11704053, 52175531); the National Key Research and Development Program of China (Grant No. 2021YFC2203600).

\section*{Disclosures}
The authors declare that there are no conflicts of interest related to this article.

\bibliography{ref}

\end{document}